\documentclass[fleqn,usenatbib]{mnras}
\usepackage{newtxtext,newtxmath}
\usepackage[T1]{fontenc}

\DeclareRobustCommand{\VAN}[3]{#2}
\let\VANthebibliography\thebibliography
\def\thebibliography{\DeclareRobustCommand{\VAN}[3]{##3}\VANthebibliography}

\usepackage{graphicx}	
\usepackage{amsmath}	
\usepackage{verbatim}   
\usepackage{xcolor}
\usepackage{graphicx}
\usepackage{threeparttable}

\title[Evolution of reverberation lags]{Tracing the evolving X-ray reverberation lags within an individual AGN light curve}

\author[N. Nakhonthong]{N. Nakhonthong$^{1}$, P. Chainakun$^{1,2}$\thanks{E-mail: \href{mailto:pchainakun@g.sut.ac.th}{pchainakun@g.sut.ac.th}}, W. Luangtip$^{3,4}$, A. J. Young$^{5}$ \\
$^1$School of Physics, Institute of Science, Suranaree University of Technology, Nakhon Ratchasima 30000, Thailand\\
$^2$Centre of Excellence in High Energy Physics and Astrophysics, Suranaree University of Technology, Nakhon Ratchasima 30000, Thailand\\
$^3$Department of Physics, Faculty of Science, Srinakharinwirot University, Bangkok 10110, Thailand\\
$^4$National Astronomical Research Institute of Thailand, Chiang Mai 50200, Thailand\\
$^5$H.H. Wills Physics Laboratory, Tyndall Avenue, Bristol BS8 1TL, UK}
\date{Accepted XXX. Received YYY; in original form ZZZ}

\pubyear{2023}

\begin{document}
\label{firstpage}
\pagerange{\pageref{firstpage}--\pageref{lastpage}}
\maketitle

\begin{abstract}

We present the Granger causality (GC) test for the X-ray reverberation analysis of Active Galactic Nuclei (AGN). If the light curves in the continuum-dominated band help predict (Granger cause) those dominated by reflection, the Granger lags that associate to the intrinsic reverberation lags can be inferred. We focus on six AGN observed by {\it XMM-Newton}, including the sources well-known to exhibit clear X-ray reverberation lags (IRAS~13224--3809 and 1H~0707--495) and those in which reverberation signatures are not well confirmed (MCG--6-30-15, IZW1, Mrk~704 and Mrk~1040). We employ the sliding-window algorithm and estimate the Granger (intrinsic) Fe-L lags along the light curve as the window moves through. This reveals the evolving lags towards the end of some individual observations, suggesting that the corona varies progressively. Occasionally, we observe two clearly separate lags that suggest an extended corona consisting of two zones while producing competing reverberation of two lags. While the GC test is purely hypothetical and might not explain true causality, our conclusion is that the lags are present and could be understood as reverberation lags. Assuming the lags changing solely with the corona, we find that the IRAS~13224--3809 corona varies between $\sim 10$--25~$r_{\rm g}$ and sometimes move to $\gtrsim 50$~$r_{\rm g}$. The corona of 1H~0707--495 and MCG--6-30-15 may be analogous to that of IRAS~13224--3809, while in IZw1, Mrk~704 and Mrk~1040 a more compact corona is expected.

\end{abstract}

\begin{keywords}
accretion, accretion discs -- black hole physics -- galaxies: active -- X-rays: galaxies
\end{keywords}

\section{Introduction}

In active galactic nuclei (AGN), the central supermassive black hole accretes large amounts of matter, releasing vast amounts of energy across the electromagnetic spectrum. The X-rays are thought to primarily originate from a cloud of hot electrons close to the black hole, called the corona, through the inverse Compton scattering of low energy (optical/UV) photons from the accretion disc with the coronal electrons \citep{Rybicki1986}. This produces the X-ray emission which is usually called the primary X-ray continuum, whose spectrum is characterized by a cut-off power-law where its slope and cut-off energy provides information about the physical processes related to the coronal properties \citep{Pozdnyakov1983}. Some X-rays from the corona can interact with the accreting gas, resulting in the reflection spectrum where  characteristic features such as photoelectric absorption and fluorescent line emission are imprinted \citep{George1991, Ross1999, Ross2005, Garcia2014}. Studying the reflection features can reveal information about the geometry and properties of the innermost region closest to the host black hole \citep[e.g.][]{Fabian2000, Fabian2003, Reynolds2003}.

AGN also exhibit large amplitude X-ray variability on various timescales. While the corona produces continuum X-rays, the subsequent reflection X-rays from different disc regions lead to the observed reverberation time delays due to their longer distances travelled to the observer. The delays of the reflection in response to continuum variations (i.e. reverberation lags) occur on short timescales for the inner disc reflection, and measuring these lags allows us to study the dynamics and structure of disc-corona system \citep{Reynolds1999, Uttley2014, Cackett2021}.

X-ray reverberation delays are typically studied using Fourier techniques, such as by analysing the lag-frequency spectrum \citep{Fabian2009, Zoghbi2010, Wilkins2013, Cackett2014, Emmanoulopoulos2014, Epitropakis2016, Caballero2018}, the lag-energy spectrum \citep{Zoghbi2013, Chainakun2016, Mastroserio2020}, or the power spectral density (PSD) profiles \citep{Papadakis2016, Emmanoulopoulos2016, Mankatwit2023}. By evaluating the X-ray reverberation lags observed in several AGN, the correlation between the amplitude of the lag and the central mass has been revealed \citep[e.g.][]{DeMarco2013, Kara2016, Hancock2022}. The evolving corona in an individual AGN such as in IRAS~13224--3809 was also evidenced \citep{Alston2020, Caballero2020, Chainakun2022b}, where during the lowest and highest luminosity state, the coronal height changed from $\sim 6~r_{\rm g}$ to $\sim 20~r_{\rm g}$ ($r_{\rm g}=GM/c^{2}$, where $M$ is the black hole mass, $G$ is the gravitational constant and $c$ is the speed of light).   

Recently, \cite{Chainakun2023} applied a Granger causality (GC) test \citep{Granger1969} to investigate the significance of the lags between the light curves extracted in the 0.3–1 keV (reflection dominated, soft) and 1.2–5 keV (continuum dominated, hard) bands from 14 {\it XMM-Newton} observations of IRAS~13224--3809. The GC is a hypothesis test used to statistically assess whether one time series is a factor that can be useful for forecasting another time series, in the way that one time series leads (or Granger-causes) the other. \cite{Chainakun2023} found that the majority of the lags from the GC test in IRAS~13224--3809 is $\sim 200$--500~s, and they could be interpreted as the intrinsic X-ray reverberation lags. The hints of the lag variability within some individual observations are also observed, suggesting the corona is evolving from the beginning towards the end of the observation. 

While the GC test provides us a unique tool to determine the intrinsic lags, the bin size of the light curves can influence the lag significance. Furthermore, a major focus in \cite{Chainakun2023} is on utilizing either the full or roughly half of the light curves to infer the lags. Here, we advance this by employing the sliding window algorithm, so that the Granger lags can be identified in a more specific temporal window within a single light curve. By moving the window along the light curve, the GC test can be examined across all segments. This allows us to probe the evolving lags continuously throughout the observation. We revisit IRAS~13224--3809 and also explore more AGN samples including 1H~0707--495, MCG--6-30-15, 1ZW1, Mrk~704 and Mrk~1040. Among these sources, IRAS~13224--3809 and 1H~0707--495 are known to display complex X-ray variability that show clear X-ray reverberation lags \citep[][and discussion therein]{Alston2020, Hancock2023,Mankatwit2023}, but the others are those that the reverberation signatures are not well confirmed \citep[e.g.][]{DeMarco2013, Kara2016, Chainakun2022a}. Therefore, this work can shed light on the various evolving lags that are suggested by the GC test in various AGN sources exhibiting different reverberation properties.

The observations and data reduction for all AGN investigated here are presented in Section 2. The concept of the GC test as well as the model assumptions and sliding window technique applied in this work are explained in Section 3. The Granger lags which could be interpreted as the intrinsic X-ray reverberation lags then are estimated directly from the AGN light curves. In Section 4, we present the theoretical models of these lags as well as the global features of variable lags seen in these AGN. We discuss and summarise the results in Section 5.

\section{Observations and data reduction}

The AGN sample analysed in this work was previously observed by {\it XMM-Newton} observatory and can be downloaded from {\it XMM-Newton Science} Archive (XSA).\footnote{\url{http://nxsa.esac.esa.int}} Here, the same observational data of the AGN IRAS~13224--3809 and 1H~0707--495 studied in recent work (i.e., \citealt{Mankatwit2023}) were also alternatively analysed by a novel method proposed in this paper, along with the additional sample of the AGN MCG--6-30-15, 1ZW1, Mrk~704 and Mrk~1040; all sample data are summarised in Table~\ref{tab:xmm_obs}. Indeed, the data reprocessing were basically the same with that performed in \citeauthor{Mankatwit2023} (\citeyear{Mankatwit2023}; see also \citealt{Chainakun2022b}). In brief, the data were reprocessed as well as removed periods which were highly affected by background flaring following the official {\it XMM-Newton} data analysis guide;\footnote{\url{https://www.cosmos.esa.int/web/xmm-newton/sas-threads}}  only pn data were used to utilise their high effective area and temporal resolution. Then, the background-subtracted light curves of each observation were extracted from the circular region with encircled energy fraction of $\sim$ 90 per cent centred on the AGN location while the background region are the nearby, source free region; these were extracted in two different wavebands -- i.e., the soft (0.3–1 keV) and hard (1.2–5 keV) energy bands -- with the timing resolution of 1 second. The obtained light curves then were represented as the observational data for analysing by method purposed in this work. Note that for the observational ID 0743050301 and 0743050801 of the AGN IZW1, each observation was not operated continuously, but was split into five segments of observation; this is because the original purpose of these observations were optimized for high-resolution spectroscopic study of outflows from the AGN (PI E. Costantini; see also \citealt{Wilkins2017}). This results in the observational data of $\sim 20$ ks for each segment, and, in this study, we analyzed each individual segmental light curve separately -- i.e., treating each segment as individual data similar to that was obtained from different observations -- to avoid the data gaps between the segments.

In addition, to allow us to compare the quality of real data with that of the simulated data, we also determined the quality of each dataset -- i.e. how much the source signal is affected by the actual background level -- by calculating the signal to noise ratio (S/N) defined as:

\begin{equation}
\text{S/N} = \frac{C_{\rm source}}{\sqrt{C_{\rm source} + C_{\rm background}}} = \frac{R_{\rm source}}{\sqrt{R_{\rm source} + R_{\rm background}}} \times \sqrt{T},
\label{eq:s/n}
\end{equation}
where $R$ is the count rate, being equivalent to count ($C$) divided by the exposure time ($T$). Here, the S/N in the hard band was calculated, and used to represent the quality of each dataset, as shown in column 6 of Table~\ref{tab:observations}.

\begin{table*}
\begin{center}
   \caption{\emph{XMM-Newton} observations of the AGN sample.} \label{tab:xmm_obs}
   \label{tab:observations}
   \begin{threeparttable}
    \begin{tabular}{lccccc}
    \hline
    Observation ID & Revolution number & ~~~~~Observed date~~~~~ & ~Exposure time$^{a}$~ & ~Count rate$^{b}$~ & ~~~~S/N$^{c}$~~~~ \\
    & & & (ks) & (cnt s$^{-1}$) & \\
    \hline
    \multicolumn{5}{c}{\bf IRAS~13224--3809} \\ 
0673580101	&	2126	&	2011-07-19	&	34.08	& 0.11 & 111\\
0673580201	&	2127	&	2011-07-21	&	49.26	& 0.09 & 110\\
0673580301	&	2129	&	2011-07-25	&	52.03	& 0.04 & 68\\
0673580401	&	2131	&	2011-07-29	&	85.11	& 0.17 & 132\\
0780561301	&	3038	&	2016-07-10	&	112.35	& 0.20 & 144\\
0780561501	&	3043	&	2016-07-20	&	93.22	& 0.08 & 94\\
0780561601	&	3044	&	2016-07-22	&	97.85	& 0.26 & 176\\
0780561701	&	3045	&	2016-07-24	&	100.13	& 0.11 & 117\\
0792180101	&	3046	&	2016-07-26	&	110.64	& 0.11 & 112\\
0792180201	&	3048	&	2016-07-30	&	110.55	& 0.14 & 122\\
0792180301	&	3049	&	2016-08-01	&	86.44	& 0.05 & 76\\
0792180401	&	3050	&	2016-08-03	&	98.63	& 0.48 & 220\\
0792180501	&	3052	&	2016-08-07	&	102.66	& 0.18 & 138\\
0792180601	&	3053	&	2016-08-09	&	101.50	& 0.48 & 231\\
\hline
\multicolumn{6}{c}{\bf 1H~0707--495} \\
0110890201	&	159	&	2000-10-21	&	29.94	& 0.09 & 52\\
0148010301	&	521	&	2002-10-13	&	57.98	& 0.44 & 162\\
0506200201  &   1361 &   2007-05-16  &   13.32   & 0.02  & 24\\
0506200301	&	1360	&	2007-05-14	&	33.38	& 0.20 & 78\\
0506200401  &   1387    &   2007-07-06  &   4.73    & 0.06  & 43\\
0506200501	&	1379	&	2007-06-20	&	20.52	& 0.44 & 122\\
0511580101	&	1491	&	2008-01-29	&	68.69	& 0.29 & 167\\
0511580201	&	1492	&	2008-01-31	&	43.88	& 0.34 & 166\\
0511580301	&	1493	&	2008-02-02	&	37.39	& 0.20 & 131\\
0511580401	&	1494	&	2008-02-04	&	50.62	& 0.22 & 133\\
0653510301	&	1971	&	2010-09-13	&	93.30	& 0.35 & 177\\
0653510401	&	1972	&	2010-09-15	&	78.46	& 0.40 & 204\\
0653510501	&	1973	&	2010-09-17	&	81.51	& 0.27 & 158\\
0653510601	&	1974	&	2010-09-19	&	79.65	& 0.30 & 175\\
0554710801	&	2032	&	2011-01-12	&	48.54	& 0.02 & 40\\
\hline
\multicolumn{6}{c}{\bf MCG--6-30-15} \\
0029740101	&	301	&	2001-07-31	&	53.16	& 5.65 & 540\\
0029740701	&	302	&	2001-08-02	&	107.21	& 8.84 & 823\\
0029740801	&	303	&	2001-08-04	&	96.47	& 7.53 & 755\\
0111570101	&	108	&	2000-07-11	&	39.16	& 4.70 & 359\\
0111570201	&	108	&	2000-07-11	&	43.92	& 6.14 & 452\\
0693781201	&	2407	&	2013-01-29	&	105.74	& 9.31 & 866\\
0693781301	&	2408	&	2013-01-31	&	114.08	& 6.21 & 707\\
0693781401	&	2409	&	2013-02-02	&	25.74	& 3.01 & 297\\
\hline
\multicolumn{6}{c}{\bf IZW1} \\
0110890301	&	464	&	2002-06-22	&	17.60	& 2.36 & 199\\
0300470101	&	1027	&	2005-07-18	&	68.84	& 1.30 & 261\\
0743050301$^{*}$	&	2768	&	2015-01-19	&	108.53	& 1.42 & 124\\
0743050801$^{*}$	&	2769	&	2015-01-21	&	89.38	& 1.36 & 142\\
0851990101	&	3680	&	2020-01-12	&	49.61	& 0.99 & 211\\
0851990201	&	3681	&	2020-01-14	&	62.03	& 1.18 & 221\\
\hline
\multicolumn{6}{c}{\bf Mrk~704} \\
0300240101 & 1074 & 2005-10-21 & 16.98 & 0.89 & 106\\
0502091601 & 1630 & 2008-11-02 & 82.88 & 2.22 & 361\\
\hline
\multicolumn{6}{c}{\bf Mrk~1040} \\
0554990101 & 1682 & 2009-02-13 & 63.22 & 3.91 & 527\\
0760530201 & 2871 & 2015-08-13 & 77.88 & 4.06 & 543\\
0760530301 & 2872 & 2015-08-15 & 63.34 & 3.57 & 496\\
     \hline
     \end{tabular}
    \begin{tablenotes}
    \item \textit{Note:} $^{a}$Summation of good time intervals (GTIs) obtained from {\it XMM-Newton} pn detector after high background flaring period was removed. $^{b}$The average, background-subtracted count rate obtained from pn detector in 1.2--5 keV (hard) energy band. $^{c}$The data S/N in the hard band defined by eq.~\ref{eq:s/n}. $^{*}$The observation was split into five individual segments which each one has the GTI of $\sim$20 ks (see text for further detail). 
    \end{tablenotes}
    \end{threeparttable}
    \end{center}
\end{table*}

\section{GC model and related algorithms}

\subsection{GC test}

The association between the time series can be studied using the the GC, by determining whether there is a relationship in which one time series precedes (or Granger-causes) the other \citep{Granger2004}. We define the soft (reflection dominated, 0.3–1 keV) and hard (continuum dominated, 1.2–5 keV) band light curves to be $s_{t}$ and $h_{t}$, respectively. Let us first assume that they are simply generated from the autoregressive process given by:
\begin{equation}
    s_t = \alpha_s + \sum_{i=1}^{q} a_{i} s_{t-i} + e_{t,s} \;,
\label{eq-s}
\end{equation}
\begin{equation}
    h_t = \alpha_h + \sum_{i=1}^{q} b_{i} h_{t-i} + e_{t,h} \;.
\label{eq-h}
\end{equation}
The summation term describes how the current observation depends on the $q$ preceding value, while $a_i$ and $b_i$ are the corresponding coefficients for the soft and hard bands, respectively. $\alpha_s$ and $\alpha_h$ are parameters that provide new information into both energy bands. The prediction errors (uncertainties or residuals) which occurred in the soft and hard bands are denoted by $e_{t,s}$ and $e_{t,h}$, respectively.

To test for the Granger causality under the X-ray reverberation framework, the $s_t$ is modified so that it also contains the past values of $h_{t}$. This can be given by a bivariate regressive process:
\begin{equation}
s_t = \gamma + \sum_{i=1}^{q} a_{i} s_{t-i} + \sum_{j=1}^{q} b_{j} h_{t-j} + \epsilon_{t,s} \;,
\label{eq-sh}
\end{equation}
where $\gamma$ is a constant and $\epsilon_{t,s}$ is the error from this estimating model. Note that the parameter $q$ is the upper limit of the lag value we want to test and some of the coefficients $a_i$ and $b_j$ can be zero. While the primary power-law component from the coronal emission is the main constituent in any of the X-ray bands, $s_{t}$ is usually more dominated by the reverberation flux than $h_{t}$ (e.g. the fraction of reflection found in $s_{t}$ is relatively large compared to what found in $h_{t}$). We then expect $s_t$ to lag behind $h_t$. In other words, $h_{t}$ should Granger-cause $s_{t}$ ($h \rightarrow s$) since the observed $s_{t}$ depends on past values of $h_{t}$.

Here, we perform $h \rightarrow s$ analysis by testing if any lag values of $h_t$ are statistically significant in the bivariate regressive model. All the lag values of $1, 2, 3, …, q$ are used to estimate the two regression equations which are eqs.~\ref{eq-s} and \ref{eq-sh}. The corresponding errors obtained in both equations for each lag value are evaluated using the GC indicator  
\begin{equation} 
         GC_{h \rightarrow s} = \log\frac{Var(e_{t,s})}{Var(\epsilon_{t,s})} \; ,
\label{eq-GC}
\end{equation}
to see if the error $\epsilon_{t,s}$ from eq.~\ref{eq-sh} is significantly smaller than the error $e_{t,s}$ from eq.~\ref{eq-s} (i.e. if the past values of $h_{t}$ help predict $s_{t}$). 

In other words, the GC is assessed by comparing the variances of the errors from both bivariate and univariate models. When the past value of $h_{t}$ does not predict $s_{t}$, $Var(e_{t,s}) \approx Var(\epsilon_{t,s})$ so $GC_{h \rightarrow s} = 0$, meaning that $h_{t}$ does not Granger-cause $s_{t}$. Contrarily, if $h_{t}$ Granger-causes $s_{t}$, so $h_{t}$ helps improve the prediction for $s_{t}$, then $Var(\epsilon_{t,s}) < Var(e_{t,s})$ and $GC_{h \rightarrow s}$ increases. 

Note that the GC test must be performed only when the light curves are stationary (i.e. having a constant mean, variance, and coherence, and no seasonal component). This is because the GC test infers the lags based on building a forecasting model (i.e. statistical testing if there are the causal linkages where the past values of $h(t)$ help predict $s(t)$ with any particular lag values). The non-stationary time series variables should not be included in a regression model, otherwise a problem known as spurious regression may arise. Therefore, most time series forecasting models assume stationary in the first place.

Here, the X-ray light curves, $s_t$ and $h_t$, are simulated from the red noise PSD (via {\tt stingray.simulator} \citep{Huppenkothen2019}) convolving with the disc response functions, which should link to the X-ray variability mechanism in AGN. The autoregressive model is used later as a forecasting model only to predict $s_t$ and $h_t$ after being transformed to stationary ones (normally, stationary series are easier to analyze by simpler forecasting models). In practice, the stationary light curve means that we minimize its components dependent on time, so that the forecasting time series model can be made. We follow the method outlined in \cite{Chainakun2023} where differencing together with the unit root test via Augmented Dickey-Fuller (ADF) method are used to ensure the stationarity. The GC tests are carried out using the {\tt grangercausalitytests} module in {\tt statsmodels.tsa.stattools} \citep{Seabold2010}. We employ the $p$-value to infer the significance of the Granger lags. Statistical hypothesis tests are arranged such that $p < 0.01$ means that the obtained lags are significant.

\subsection{Sliding window}

A sliding window is used to perform the GC test on different segments of the light curve. The window moves through the light curve, from left to right, allowing the Granger lags to be estimated for each segment. The window size determines the length of the segment considered at each specific time, whereas the moving step-size determines the amount of shift applied to the sliding window as it progresses each step. The choice of step size (e.g. overlapping or non-overlapping) depends on several factors and may affect the computational resolution and accuracy of the specific problem.

\begin{figure}
     \centering
    {\includegraphics[width=1.0\linewidth]{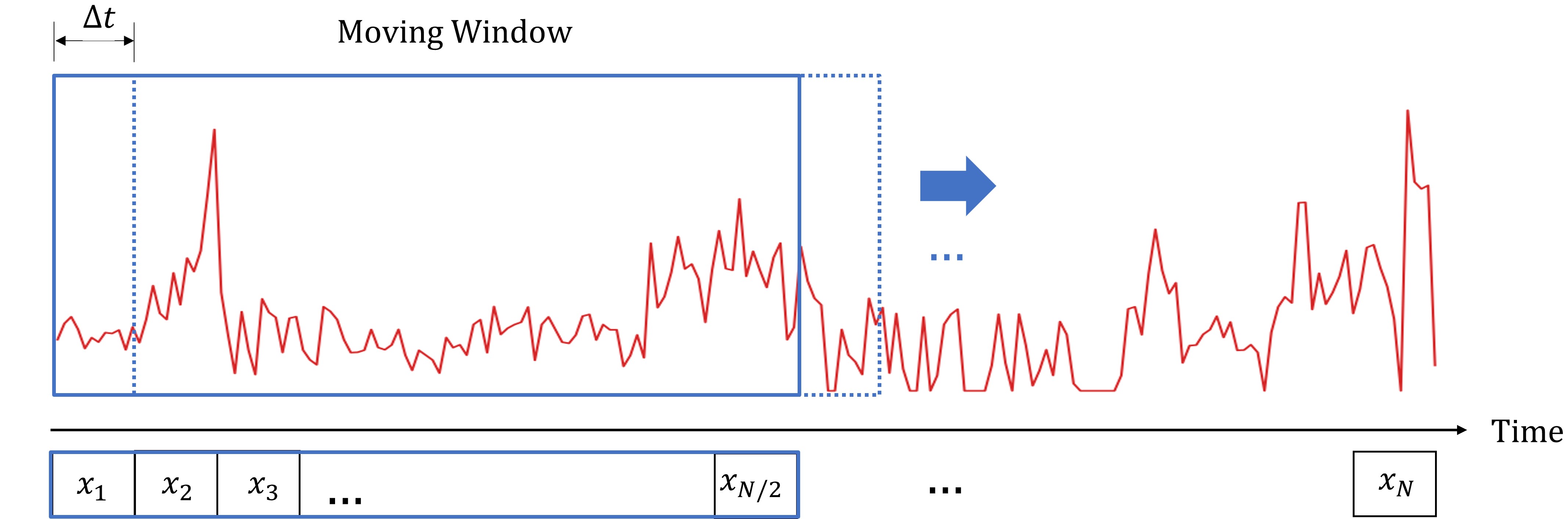}\label{sdw-b}}
    \vspace{-0.3cm}
     \caption{Sketch of moving window with a step size equal to the bin size of the light curve, $\Delta t$. The size of the window is fixed at $N\Delta t /2$.}
     \label{sd-window}
\end{figure}

To avoid redundant computations, we define the window size to be a fixed value equal to the half length of the light curve, as illustrated in Fig.~\ref{sd-window}. If the light curve contains $N$ time bins, each of which has the size of $\Delta t$, the window size is then $N\Delta t /2$. When the Granger lags for the current window have been analyzed, the window is moved by one position to the right, with the moving step equal to the bin size $\Delta t$. We do not consider the stationarity for an entire light curve or for all segments simultaneously, but instead look at the GC lag for each individual segment where the sliding-window moves through one by one segment (i.e. the mean, variance and autocovariance should be constant within each segment covered by the window, but different segments can have different statistical properties). Therefore, the estimated lags between two light curves can be varied across all segments, and stacking them together can reveal how time lags evolve continuously along the light curve within a single {\it XMM-Newton} orbit.

\subsection{Error estimates}

For the GC test in the previous section, we obtain the Granger lags of each observation with the condition of the hypothesis test (e.g. $p$-values). Different significant lag-values can be found based on different choices of binning. The error of the lag may be estimated by, for example, dividing the light curve for each window further into several sub-segments. Then, the lag is given by the average of these sub-segments, and the standard error of the mean lag is quoted as an error. This method is similar to the one used for the cross-correlation function which does not always work especially when the light-curve segments are short. Therefore, it might not be obvious how to estimate the error. For consistency, we use the time-binning size as the approximate error for the Granger lags. For example, if the time bin is 100~s (i.e. the Granger tests are performed with the time step of 100~s with $q = 1$, 2, 3, …), the obtained lags can possibly be 100~s, 200~s, 300~s, … with the estimated error of $+/-$~100~s.

\section{Results}

\subsection{Identifying the lags in the light curve}
To illustrate the Granger-lag profile used in our analysis, we construct 50~ks light curves with a binning of 1~s from the primary variation $a_t$ generated by the red noise processes in {\tt stingray.simulator} \citep{Huppenkothen2019}. The reverberation response is modelled, for simplicity, using the top-hat response function $TH_{\rm rev}$ whose profile is controlled by the centroid and the width which are set to be $\tau = 600$~s and $t_{w} = 50$~s, respectively. On the other hand, we include the response due to the disc propagating fluctuations that operate on relatively long timescales in the hard band using the top-hat function $TH_{\rm prop}$, with $\tau = 1000$~s and $t_{w} = 200$~s. The area under these responses are normalized to 1. The soft reflection-dominated and hard continuum-dominated band light curves are computed by 
\begin{equation}
s_t = a_t + R_s a_t \otimes TH_{\rm rev} + N_{s,t}\;,  \label{eq-s-rev}
\end{equation}
\begin{equation}
    h_t = a_t + R_h a_t \otimes TH_{\rm prop} + N_{h,t} \;,
\label{eq-h-rev}
\end{equation}
where $R_s$ and $R_h$ represent the fraction of reverberation and disc-propagating fluctuations contributed in the soft and hard bands. We introduce the effects of random noise to the light curves by adding uncorrelated variability with random mean and the standard deviation using {\tt numpy.random.normal}. The signal-to-noise ratio (S/N) is estimated from the mean count rate over the average error in the way that enables us to compare it with the S/N of the real data (eq.~\ref{eq:s/n}). The light curves are binned with the bin size of $\Delta t = 50$~s. Note that the bin size will affect the significance of the lags. If the bin size is chosen to be comparable to the width of the response, the characteristic response profile will become close to the delta-function and the Granger lags can be interpreted as the average reverberation time determined by the centroid of the top hat response. In reality, we do not know in advance the width of the reverberation response and binning the data too large may remove information about the width of the response. In our current approach, the variable bin-size is applied to investigate all possible significant lags.

We perform the $h \rightarrow s$ test and analyze the lags following the method outlined in \cite{Chainakun2023}. Briefly speaking, we investigate all lag values $q$ and the corresponding errors obtained from the two regression equations (eqs.~\ref{eq-s} and \ref{eq-sh}). The $p$-value is used to justify if the error from eq.~\ref{eq-sh} (where $s_t$ also contains the past values of $h_t$) is significantly smaller than that from eq.~\ref{eq-s} (where $s_t$ is explained by its own past values). We plot each lag value $q$ in terms of the corresponding $p$-values in order to infer their significance.

\begin{figure*}
    \centerline{
        \includegraphics[width=1.0\textwidth]{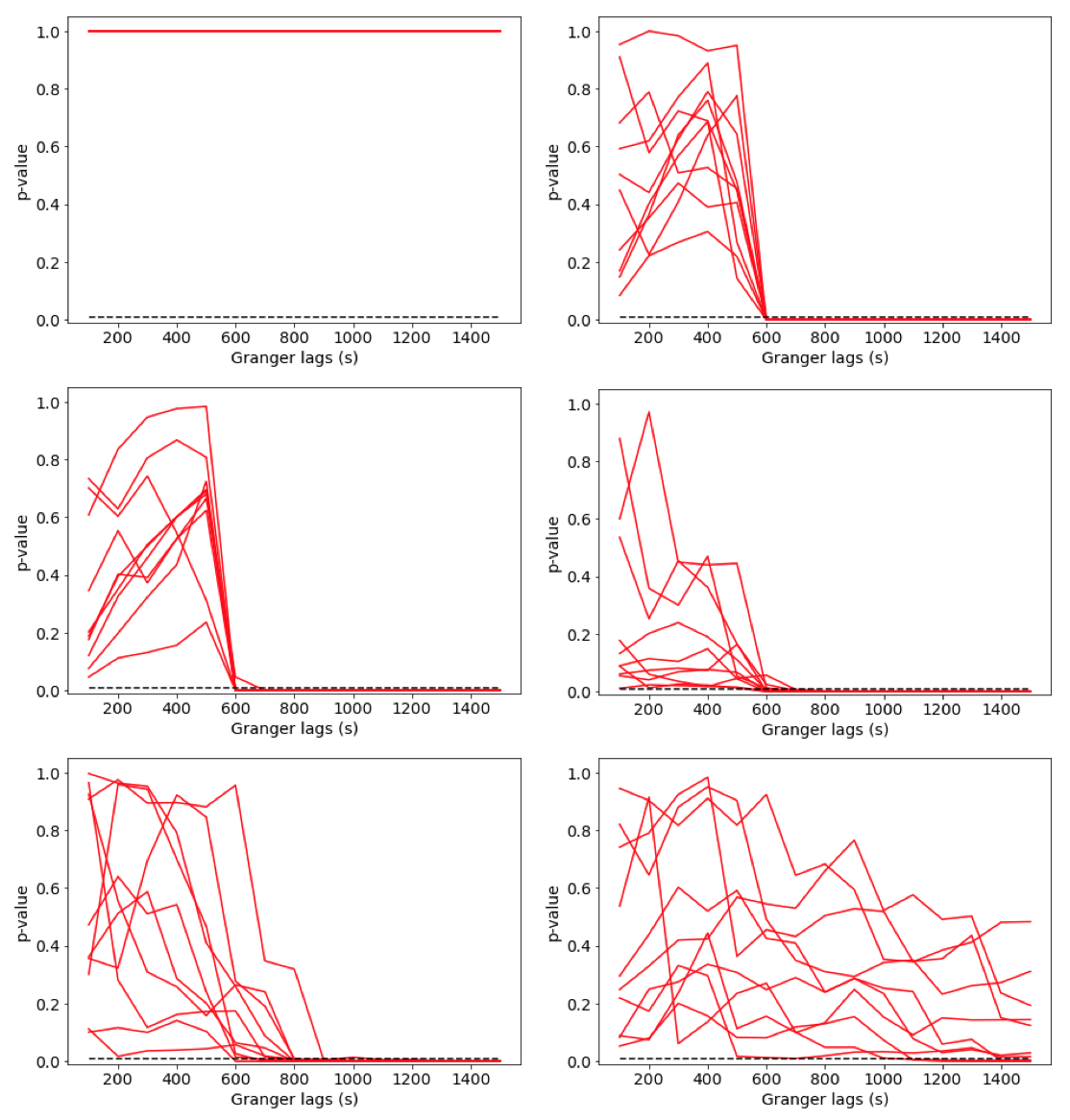}
        \put(-52,510){S/N = 192.6} 
        \put(-306,332){S/N = 181.4} 
        \put(-52,332){S/N = 151.5} 
        \put(-306,156){S/N = 110.1} 
        \put(-52,156){S/N = 83.9} 
    }

    \caption{Granger-lag profiles ($h \rightarrow s$) from 10 simulated pairs of light curves plotted in terms of the $p$-value when the light curves are identical (top-left panel) and when the top-hat response representing reverberation and propagation lags is included in the soft and hard bands, respectively. We also include uncorrelated variability to show how the estimated lags depend on the S/N of the light curves. For the reverberation response, we fix the centroid of the response time to be 600~s. The reflection fraction is fixed at $R=1$. The significance level of $p \leq 0.01$ is shown in the horizontal dashed line. The intrinsic lags of 600~s can be estimated using the minimum lag-value at the point where the curve crosses the significance level line. It can be seen that the error in estimating the lags increases with decreasing the S/N ratio, and the preferred S/N in this method is approximately more than 100.}
    \label{fig-GC-lags}
\end{figure*}
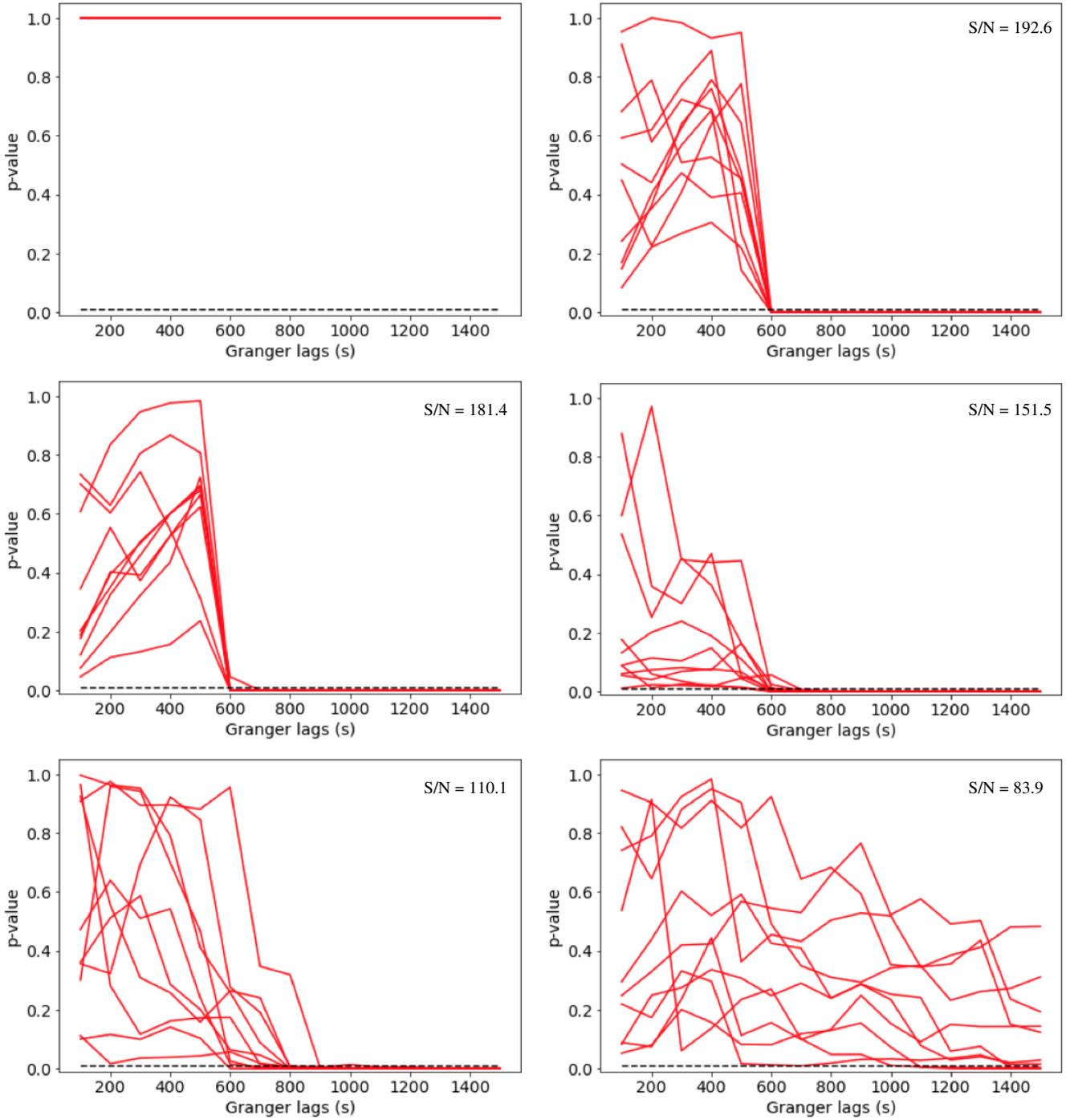

Fig.~\ref{fig-GC-lags} shows examples of the Granger-lag profiles obtained from simulated 10 pairs of the light curves. It can be seen that the reverberation-lag amplitude ($\tau = 600$~s) could be estimated from the minimum Granger-lag at $p \leq 0.01$, especially when the S/N $\gtrsim 100$. The inferred reverberation lag is also an intrinsic one since the dilution can only make the lags become less significant but does not alter the value of the lags. Note that the $p$-value can stay smaller than 0.01 even at longer values of the targeted lag (i.e. at > 600 s here). This is because the Granger causality is a statistical test to determine whether “at least” one of the lags of $h_t$ up to that particular value is significant to explain $s_t$. In other words, when at least one of the lags of $< 600$~s of $h_t$ can help predict $s_t$, it is also true that at least one of the lags of $< 800$~s can help explain the data. This is why the $p$-values for long Granger lags can be $< 0.01$, and why we should use the minimum Granger lag that is first significant as a proxy of the reverberation lags. However, the two lags can still be probed using variable bin sizes, such as in Fig.~6 of \cite{Chainakun2023}. The random noise and effects of the competing process such as the disc propagating fluctuations that operate on much longer timescales affects more on the significance of the lag than on manifesting the lag values \citep[see also Appendix A and B in][]{Chainakun2023}.

\subsection{Identifying the lags using the sliding window}

\begin{figure}
    \centerline{
        \includegraphics[width=0.5\textwidth]{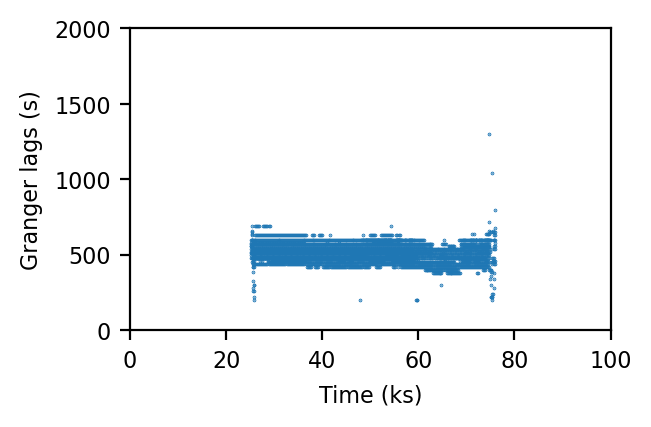}
        \put(-191,140){$~\tau = 500$~s}
}
     \vspace{-0.2cm}
    \centerline{
        \includegraphics[width=0.5\textwidth]{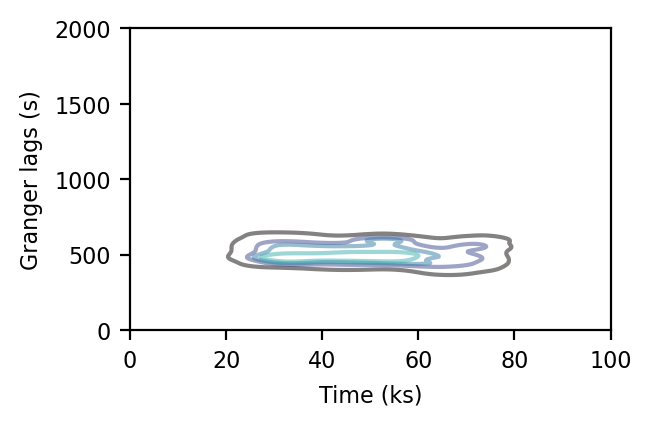}
        \put(-62,140){\textcolor{white}{$C_{\rm thres}=0$}}
}
    \vspace{-0.2cm}
    \centerline{
        \includegraphics[width=0.5\textwidth]{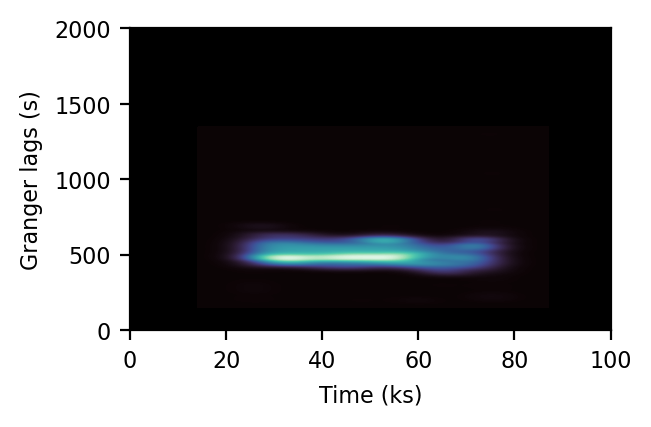}
}
    \caption{Granger-lag values detected via the sliding window along the simulated light curve of 100~ks when the disc reverberation response assumed to be a top-hat function ($TH_{\rm rev}$) with the centroid response time of $\tau =500$~s (top panel). The corresponding contour (middle panel) and density (bottom panel) maps of the distribution of the Granger lags estimated using the KDE are also presented. The density units are arbitrary where higher density distribution of the lags is shown with lighter colors.}
    \label{fig-1process}
\end{figure}

Rather than considering an entire light curve to have a single lag value, the sliding window is applied to estimate the lags along the light curve. The Granger-lags along the light curves are probed using several bin sizes, so that the results can be compared and discussed. The bin sizes used in this illustration start from the smallest size of $\Delta t_{i}=100$~s and increase with the step size given by $\delta t_{\rm step}=10$~s until it reaches the maximum size of $\Delta t_{f}=500$~s. All lags found when the window moves through the sequence are recorded and plotted against the the mid-time of the sliding window. We simulate the light curve as before and assume the $TH_{\rm rev}$ of $\tau = 500$~s in the soft band. The result is shown in Fig.~\ref{fig-1process}. It can be seen that the almost constant Granger-lags of $\sim 500$~s can be detected along the segments where the window moves through. We employ the Kernel Density Estimation (KDE) function available in {\tt seaborn} \citep{Waskom2021} to estimate the density of the population of the obtained lags with a Gaussian kernel\footnote{It is a limitation of the current GC approach that the lags should be tested across different bin sizes, and ones can see all possible lags obtained when bin size is varied in the KDE plot}. The middle and bottom panels of Fig.~\ref{fig-1process} represents the contour lines depicting the density levels of the lag distribution and the corresponding smoothed version of the density, respectively. It is normalized such that all the probabilities for all the lag values found in this map add up to 1, meaning 100 per cent of the lag falls under the distribution plot. The units of the density are arbitrary where lighter colors show higher density values for the obtained lag distribution. The almost constant lags that cluster around $\sim 500$~s is analogous to, e.g., the reverberation lags from the lamp-post corona staying at a fixed height on the symmetry axis of the black hole. Note that we also investigate the effects of the chosen bin size in Appendix~\ref{sec:a1}).

\begin{figure}
    \centerline{
        \includegraphics[width=0.5\textwidth]{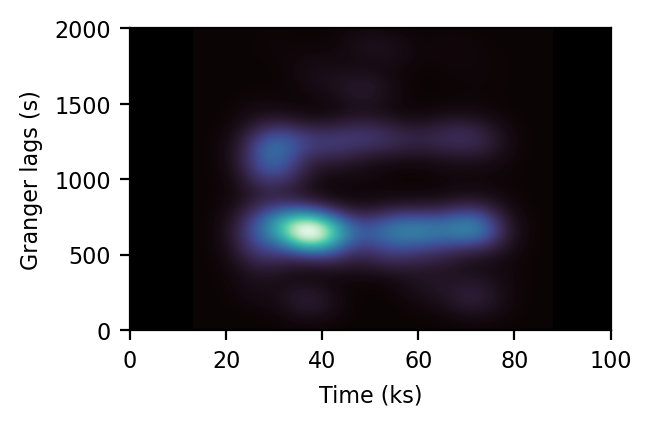}
        \put(-191,140){\textcolor{white}{$~\tau_{1} = 600$~s, $\tau_{2} = 1200$~s}}
        \put(-65,140){\textcolor{white}{$R_{s,1} = 0.6$}}
        \put(-65,128){\textcolor{white}{$R_{s,2} = 0.3$}}
}
     \vspace{-0.2cm}
    \centerline{
        \includegraphics[width=0.5\textwidth]{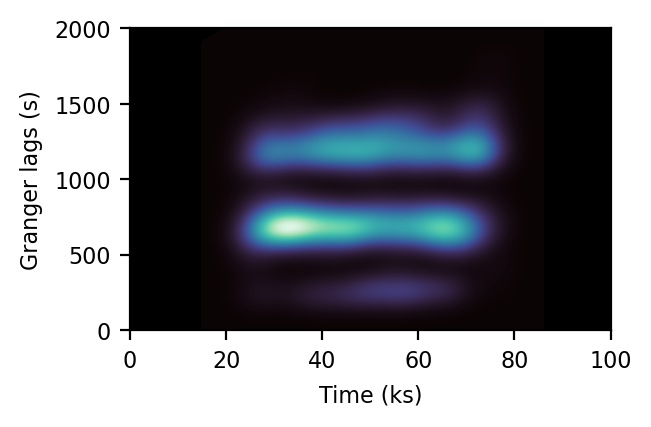}
        \put(-65,140){\textcolor{white}{$R_{s,1} = 0.6$}}
        \put(-65,128){\textcolor{white}{$R_{s,2} = 0.4$}}
}
    \caption{Same as in the bottom panel of Fig.~\ref{fig-1process}, but when the soft band contains two top-hat response functions of $\tau = 600$~s and 1200~s representing a dual lamp-post case. While the fraction of the reflected flux of the lower source is fixed at $R_{s,1}=0.6$, that of the upper source is varied to be $R_{s,2}=0.3$ (top panel) and 0.4 (bottom panel). A dual lamp-post X-ray corona can produce two separate lags appearing in the Granger-lag profiles.}
    \label{fig-2process}
\end{figure}
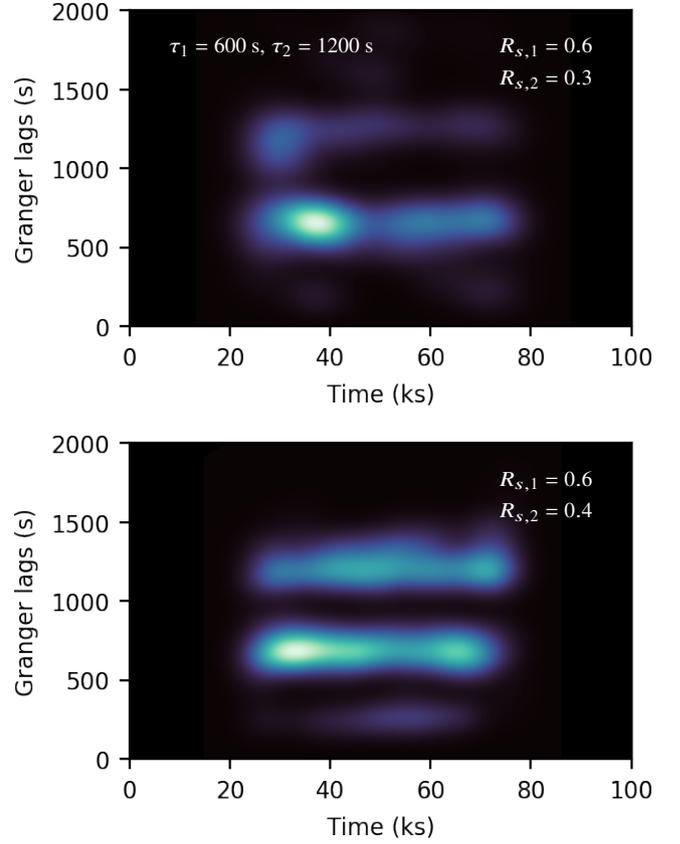

Now, we investigate a dual lamp-post case intended to explain the vertical extent of the corona by adding a second top-hat function ($TH_{\rm rev,2}$) mimicking the reverberation response due to a second (upper) lamp-post source \citep{Chainakun2017, Lucchini2023}. The soft reflection-dominated band light curve is then given by 
\begin{equation}
s_t = a_t + a_t \otimes (R_{s,1}TH_{\rm rev,1} + R_{s,2}TH_{\rm rev,2}) + N_{s,t}\;, \label{eq-s-rev2}
\end{equation}
where $R_{s,1}$ and $R_{s,2}$ regulate the importance of the reflected flux between the lower and the upper source, respectively. For the lower source, there is a larger amount of photons incident the disc so $R_{s,1} > R_{s,2}$. We simulate the light curve using $\tau_1=600$~s and $t_{w,1}=80$~s for $TH_{\rm rev,1}$, and $\tau_2=1200$~s and $t_{w,2}=50$~s for $TH_{\rm rev,2}$. The obtained Granger-lag profiles are presented in Fig.~\ref{fig-2process}. The results demonstrate that a dual lamp-post corona can produce the competing lags in two separate zones specifically to the lags due to each X-ray source. Note that the width of the response as well as the factor $R_{s,1}$ and $R_{s,2}$ affect the significance of the lags, but the lags can be approximated as the intrinsic reverberation lags with amplitudes given by $\tau_1$ and $\tau_2$. Smaller $R_{s,2}$ does not decrease the lag amplitude but it can lead to the less significance of the detected time lags as in Fig.~\ref{fig-2process}.

\begin{figure*}
    \centerline{
        \includegraphics[width=0.9\textwidth]{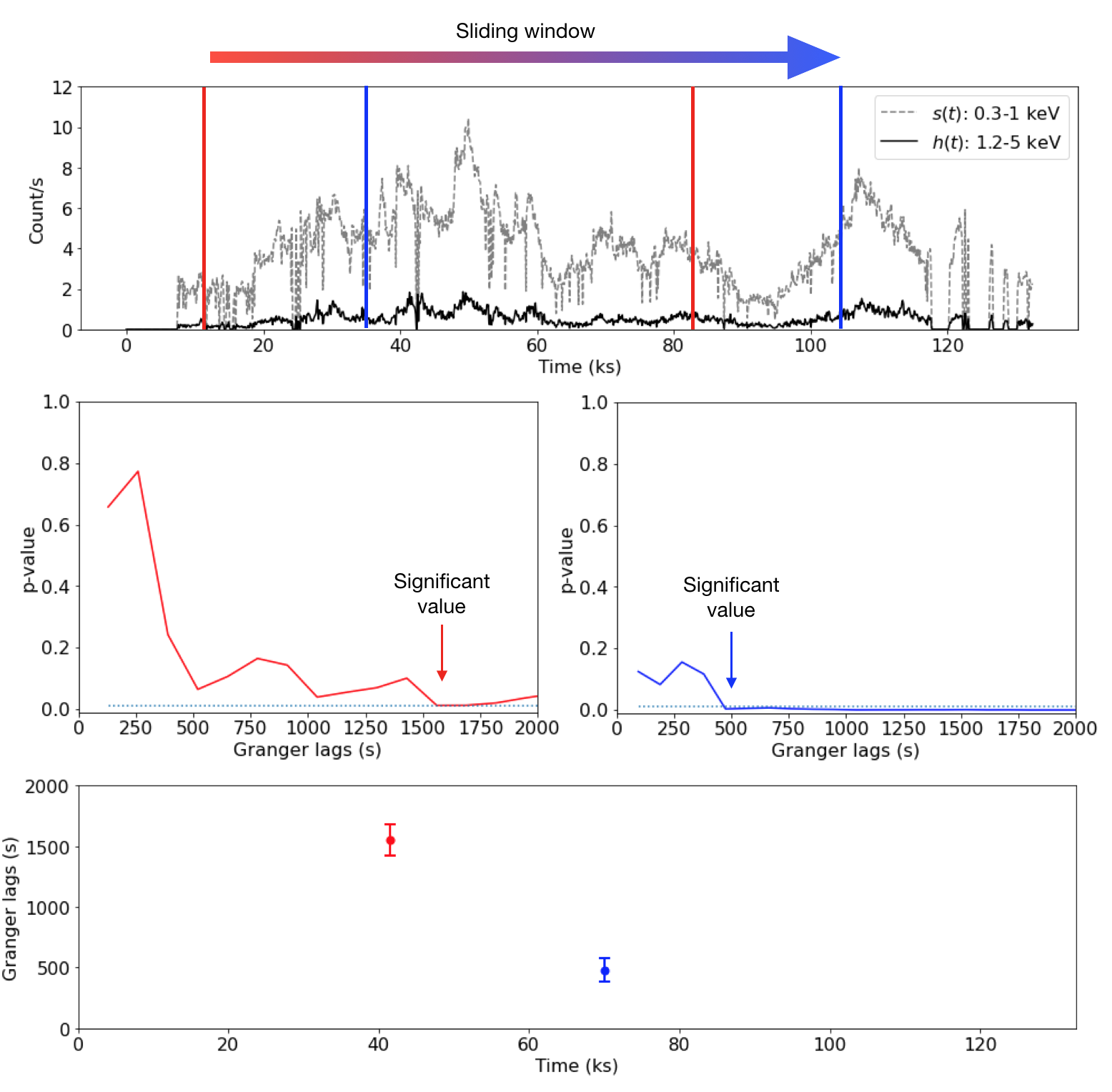}
    }
    \caption{Illustration of the sliding window used to process and analyze the Granger-lags from each sequence of the light curve of IRAS~13224--3809 rev. no. 3053 where the window of fixed size moves through (top panel). In this example, the lags ($p < 0.01$) from two specific sequences covering by red and blue windows are found to be $\sim 1570$~s to $\sim 700$~s, respectively (middle panels). The obtained lags are plotted against the light-curve time given by the mid-time of the corresponding sliding window (bottom panel). While the time-bin size is used as the approximate error for the lags, we note that the corresponding KDE plot (as shown in Fig.~\ref{fig-iras3053-map}) should also provide the idea of the distribution of all possible lags estimated with this method. 
    \label{fig-identify}}
\end{figure*}

Fig.~\ref{fig-identify} shows examples of the Granger-lag profiles ($h \rightarrow s$) derived from two different segments of IRAS~13224--3809 light curve rev. no. 3053. We can see that the hard continuum-dominated band Granger-causes (leads) the soft reflection-dominated band at particular lag values, consistent with the reverberation framework where we expected to see the soft band lagging the hard band. The mid-time of the sliding window is used to record the lag value seen in that segment. When the mid-time of the sliding window changes from 49~ks to 70~ks, the lags are found to decrease from $\sim 1570$~s to $\sim 700$~s. 

\begin{figure}
    \centerline{
        \includegraphics[width=0.5\textwidth]{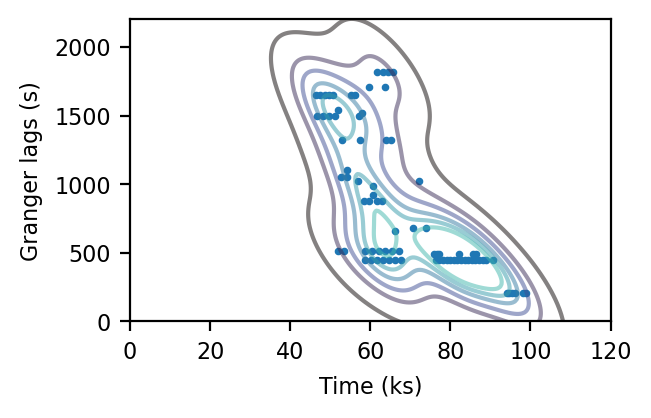}
        \put(-82,140){IRAS~13224--3809}
        \put(-70,130){(rev. no. 3053)}
}
     \vspace{-0.2cm}
    \centerline{
        \includegraphics[width=0.5\textwidth]{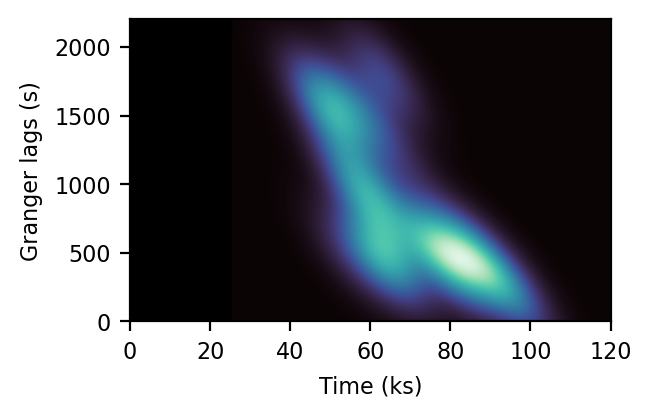}
}

    \vspace{-0.2cm}
    \caption{Top panel: Granger-lag values (data points) detected along the light curve of IRAS~13224--3809 rev. no. 3053 and the contour lines showing the density distribution of the data. Bottom panel: Corresponding smoothed density distribution of the lags where lighter colors show higher density values. We can see the tendency of the lags that decrease towards the end of the observation.}
    \label{fig-iras3053-map}
\end{figure}

By applying the sliding window and binning the data using various sizes of the time bins, we can identify the lags continuously along the light curve as well as the distribution of the significant lags dependent on the choice of the time bins. Fig.~\ref{fig-iras3053-map} shows all Granger-lag values detected along the light curve of IRAS~13224--3809 rev. no. 3053 once the sliding window is applied. We use the KDE to highlight how the distribution of the lags is found and how the lags change across the light curve. Here, the trend of decreasing lags towards the end of the light curve is evidenced. Next section, we explore the general trends of these lags for all observations of our targeted AGN.

\subsection{Global features of variable lags}

\begin{figure*}
    \centerline{
        \includegraphics[width=1.0\textwidth]{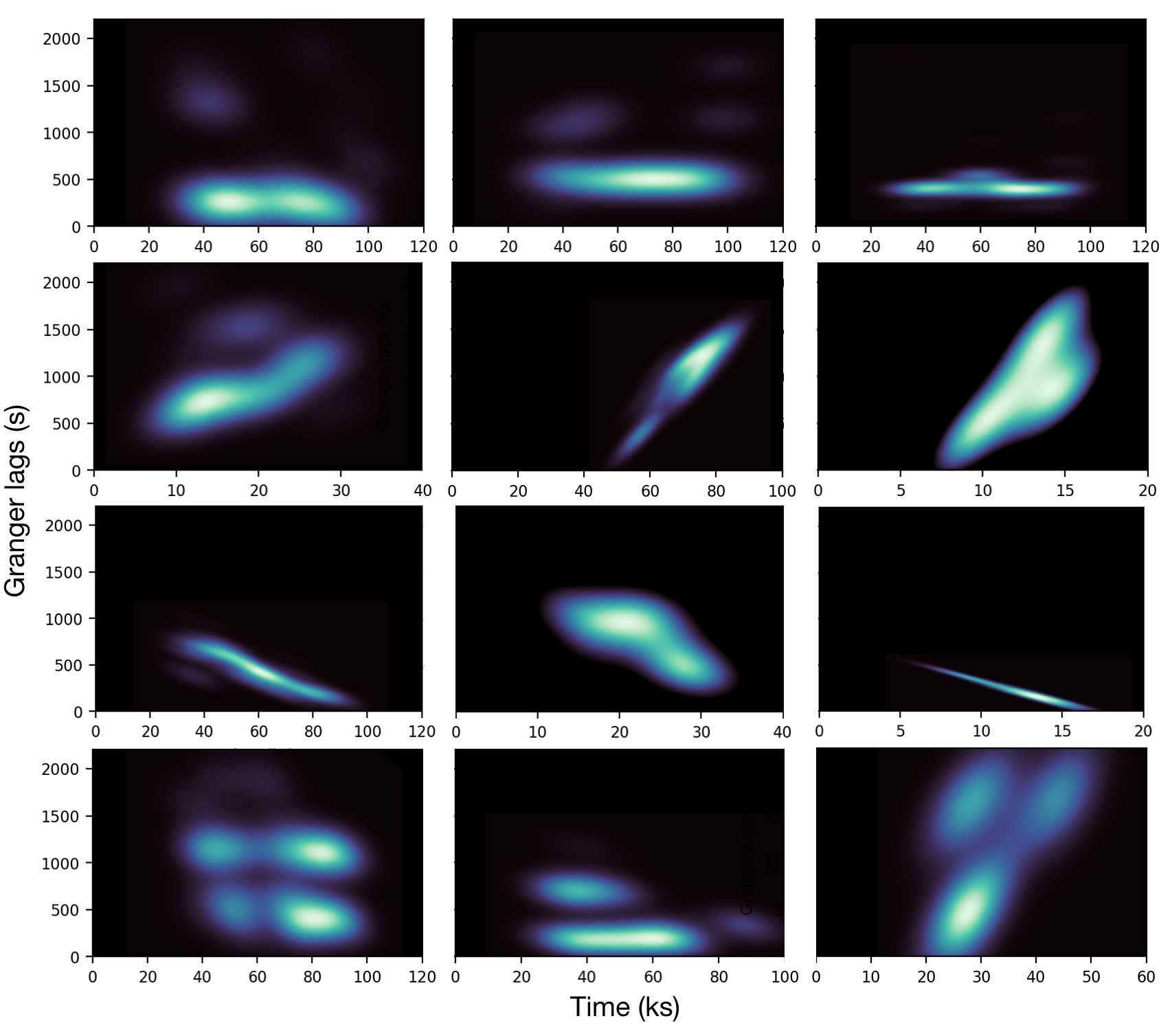}
        \put(-460,425){\textcolor{white}{IRAS~13224--3809}}
        \put(-460,415){\textcolor{white}{ (rev. no. 3043)}}
        
        \put(-302,425){\textcolor{white}{IRAS~13224--3809}}
        \put(-302,415){\textcolor{white}{ (rev. no. 3044)}}

        \put(-145,425){\textcolor{white}{1H~0707--495}}
        \put(-145,415){\textcolor{white}{ (rev. no. 1972)}}
        \put(-460,317){\textcolor{white}{1H~0707--495}}
        \put(-460,307){\textcolor{white}{(rev. no. 159)}}
 
        \put(-302,317){\textcolor{white}{MCG--6-30-15}}
        \put(-302,307){\textcolor{white}{ (rev. no. 2407)}}

        \put(-145,317){\textcolor{white}{Mrk~704}}
        \put(-145,307){\textcolor{white}{ (rev. no. 1074)}}

        \put(-460,210){\textcolor{white}{IRAS~13224--3809}}
        \put(-460,200){\textcolor{white}{(rev. no. 3052)}}
        
        \put(-302,210){\textcolor{white}{1H~0707--495}}
        \put(-302,200){\textcolor{white}{ (rev. no. 1387)}}

        \put(-145,210){\textcolor{white}{1Zw1}}
        \put(-145,200){\textcolor{white}{ (rev. no. 2769-4)}}

        \put(-460,103){\textcolor{white}{IRAS~13224--3809}}
        \put(-460,93){\textcolor{white}{(rev. no. 2127)}}
        
        \put(-302,103){\textcolor{white}{1H~0707--495}}
        \put(-302,93){\textcolor{white}{ (rev. no. 1491)}}

        \put(-145,103){\textcolor{white}{Mrk~704}}
        \put(-145,93){\textcolor{white}{ (rev. no. 1630)}}
    
    }
    \caption{General trends of the distribution of the Granger-lags found as seen in our light curves. The panels in the first, second, third and fourth rolls represent the (almost) constant, increasing, decreasing and irregular trends, respectively. Note that they are selected and catalogued by eyes. All sources presented here have the S/N $\gtrsim 100$ except IRAS~13224--3809 rev. no. 3043 (S/N = 94), and 1H~0707--495 rev. no. 159 (S/N = 52) and 1387 (S/N = 43). Higher density values for the lag distribution are plotted in lighter colors. See text for more details.} 
    \label{fig-global}
\end{figure*}

General trends of the Granger-lags seen in the light curves can be summarized as an example in Fig.~\ref{fig-global}. The lags in some observations of these AGN remain almost constant as time evolves. These include, e.g., IRAS~13224--3809 rev. nos. 3043 and 3044, and 1H~0707--495 rev. no. 1972. Their corona then should be located at an almost constant height above the black hole during the observation. Meanwhile, some observations show increasing and decreasing trends of the lags towards the end of the observation, analogous to the lags from the dynamic corona changing its height. 

The irregular trends of the variable lags are also seen in a number of particular observations. For example, in IRAS~13224--3809 rev. no. 2127, we see two separate lags at $\sim 500$~s and $\sim 1200$~s. Perhaps this suggests the corona is substantially extended, clustering in two zones unfolded before us as the competing reverberation of two lags. Interestingly, in the case of Mrk~704 rev. no. 1630, the increasing lags are observed but we see the hint of separate, competing lags that may suggest the extended corona.

\subsection{Evolving lags in 6 AGN}

\begin{figure}
    \centerline{
        \includegraphics[width=0.5\textwidth]{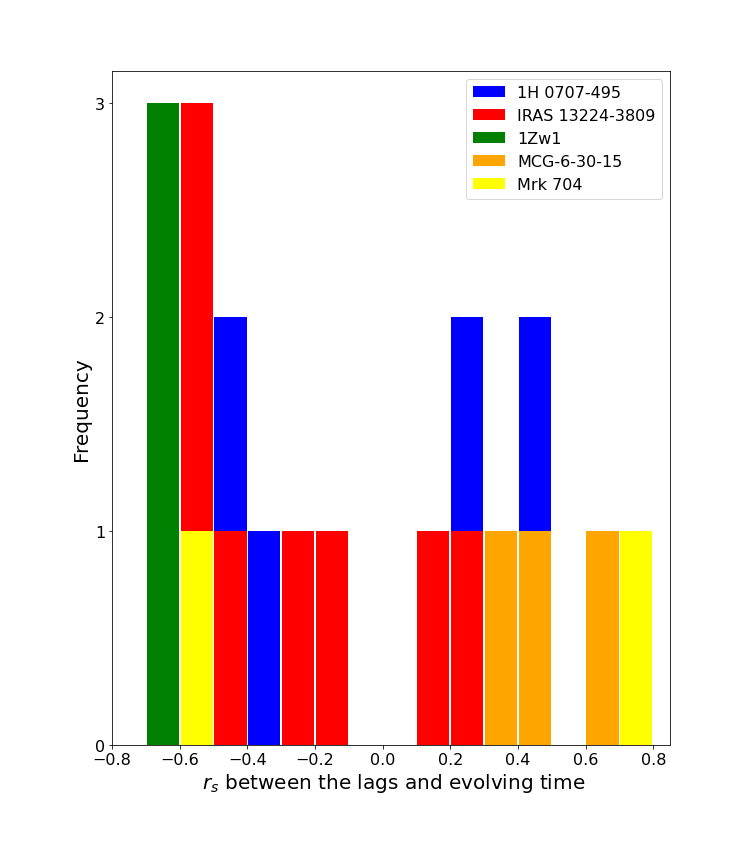}
    }
    \vspace{-0.5cm}
    \caption{Distribution of the obtained Spearman's rank correlation coefficient ($r_s$) for the correlation between the Granger-lags and evolving time. Only the samples that have significant correlation ($p<0.01$) are included. The significant GC lags correlated with the light curve time can be found in all AGN except Mrk~1040. $r_s > 0$ means the lags increase towards the end of the observation, and vice versa.
    \label{fig-rs-lag-time}}
\end{figure}

From the visual representations of the data, we attempt to determine the trend of the evolving lags by considering the Spearman's rank correlation coefficient ($r_s$). We analyze whether the Granger-lag values generally increase (or decrease) as time progresses for each individual light curve. Fig.~\ref{fig-rs-lag-time} shows the obtained monotonic correlations between the lags and the light curve time for all AGN investigated here. Only the results with $p<0.01$ are presented. The correlations are significant in 1H~0707--495, IRAS~13224--3809, 1Zw1, MCG--6-30-15, and Mrk 704. When $r_s > 0$ (or $r_s < 0$), the lags significantly increase (or decrease) towards the end of the observation. We find that the lags significantly increase (decrease) with time in 8 (or 11) observations of these AGN.

\section{Discussion and conclusion}

A traditional way to study the X-ray reverberation delays is to analyze them in the Fourier-frequency domain by calculating, e.g., the lag-frequency spectrum \citep{Fabian2009, Zoghbi2010, Wilkins2013, Cackett2014, Chainakun2015, Caballero2018} where the use of the full light curve is usually necessary. In fact, a dynamic of the X-ray corona was robustly constrained in, e.g., IRAS~13224--3809 by considering the lag observed in each of the full light curves from 16 {\it XMM-Newton} observations \citep{Alston2020,Caballero2020, Chainakun2022b}. However, analysing the lags from the full light curve can relegate them to the average value, hence preventing us from studying how the lags evolve throughout the individual observation. This work builds on the GC analysis on the intrinsic X-ray reverberation reported by \cite{Chainakun2023} where the hint of the coronal evolution towards the end of some individual observations of IRAS~13224--3809 was suggested. 

Here, the study is expanded to six AGN and the sliding window is applied to probe the variable lags along the light curve. We attempt to determine and visualize the trends of the evolving lags by considering the distribution of the obtained lags via the KDE \citep{Waskom2021}. The lags seen along the individual light curve where the sliding window moves through are likely variable (Fig.~\ref{fig-global}). This suggests a presence of the dynamical corona in the individual observations of these AGN. The simplest way to induce such intrinsic-lag variation is that the corona changes its height along the rotational axis of the black hole, as explained by the light bending model \citep{Miniutti2004}.  

From 14 {\it XMM-Newton} observations of IRAS~13224--3809, \cite{Chainakun2023} found the Granger-lags are significant in 12 observations while the evolving lags are seen only in 4 observations. By adopting variable bin sizes and applying the sliding-window technique, we find the variable and significant lags in all 14 IRAS~13224--3809 observations. Furthermore, we find the tendency of the lags to significantly increase (or decrease) towards the end of the observation in 8 (or 11) light curves from six AGN (Fig.~\ref{fig-rs-lag-time}). When this trend is not observed, the lags tend to concentrate around a specific value suggesting a corona which is less variable in size and location but, perhaps, still moving up or down along the rotational axis. Nevertheless, for the majority of the cases, it is very likely that the scatter of the obtained lags are intrinsic to the extended corona. This is because the distributions of the lags revealed to us look very different from one observation to another and may cluster in more than one zone. This can be inferred as having in some sense a competing reverberation process due to different coronal parts. 

While the exact shape of the corona is under debate, it may vary among different AGN or different observations. According to Fig.~\ref{fig-global} (bottom left panel), we see in IRAS~13224--3809 rev. no. 2127 two almost-persistent but separate lags at $\sim 500$~s and $\sim 1200$~s along the light curve. These features may be captured by a vertically extended corona simplified using two lamp-post sources as demonstrated in Fig.~\ref{fig-2process}. Therefore, this may simply be explained by dual lamp-post model \citep{Chainakun2017, Hancock2023, Lucchini2023} where there are two X-ray blobs inducing two separate reverberation lags. In terms of the GC test, the two lags from dual lamp-post source are not average as one, but instead show that both lag values can provide useful information of reverberation that statistically help explain the light curve. A more complex and extended corona \citep[e.g.][]{Wilkins2016, Chainakun2019} may be required to explain a more complex trend of the evolving lags.

The IRAS~13224--3809 corona constrained using different methods, such as the lag-frequency spectra \citep{Alston2020}, 
the combined spectral-timing model \cite{Caballero2020}, the power spectral density analysis \citep{Chainakun2022b} and the machine learning approach via the random forest regressor \citep{Mankatwit2023}, was quite consistent in that it was found to vary between $\sim 5-20~r_{\rm g}$. A study on IRAS~13224--3809 using a dual lamp-post model also suggested the corona could extend up to $\sim 20~r_{\rm g}$ \citep{Hancock2023}. Here, the intrinsic lags of $\sim 200$--500~s are found in all IRAS~13224--3809 observations, and sometimes the lags can be as large as $>1000$~s in particular segments of each individual light curve. In some observations the lags may stay almost constant at $< 500$~s towards the end of the observation (e.g. rev. nos. 3043 and 3044 in top panels of Fig.~\ref{fig-global}). Assuming a face-on disc and the central mass of $2 \times 10^{6}~M_{\odot}$ \citep{Alston2020}, we can convert the intrinsic lags directly to the true light-travel distance. The inferred distance suggests that the corona spends most of its time at $\sim10$--25~$r_{\rm g}$ and may occasionally move up to $\gtrsim 50$~$r_{\rm g}$, consistent with \cite{Chainakun2023}. The disc density as well as the X-ray luminosity that determines the ionization of the disc may affect the dilution of the lags, but they should not significantly affect the intrinsic lags that are probed by the Granger test as long as the geometry is the same. However, the Granger lags interpreted here in the context of the height-changing corona is a rough approximation since the disc can have a geometric thickness that varies among different AGN \citep{Taylor2018}. It is still possible that the changes in Granger lags observed here may be the results of the combined effects due to the changes in both corona and disc geometry.

For 1H~0707--495, the central mass can possibly be $\sim 2.3-10 \times 10^{6}~M_{\odot}$ \citep[e.g.][]{Zhou2005, Zoghbi2011, Pan2016} while the observed Granger-lags are in the same range of those seen in IRAS~13224--3809. With the larger mass and the comparable intrinsic lags, the coronal height in 1H~0707--495 should be smaller than that in IRAS~13224--3809. \cite{Wilkins2012} investigated the emissivity profiles using various coronal geometries and suggested a presence of the 1H~0707--495 corona at $\sim 2~r_{\rm g}$ extending radially outwards to $\sim 30~r_{\rm g}$. Recent studies on the relativistic reflection spectra of 1H~0707--495 suggested that its corona should be very compact and located at $h \sim 3~r_{\rm g}$ with an extended size of $\lesssim 1~r_{\rm g}$ \citep{Szanecki2020}. A proposed corona at $\sim 3$--$4~r_{\rm g}$ was also consistent with \cite{Dauser2012} and \cite{Caballero2018}. However, the coronal extent must be large enough to receive the seed photons from the accretion disc in order to produce the continuum X-rays. Regarding this, \cite{Dovciak2016} proposed that the 1H~0707--495 corona may locate at $\sim 2-5~r_{\rm g}$ and its size may extend outwards to $\sim 20~r_{\rm g}$ during its maximum flux. Recently, \cite{Hancock2023} showed that the frequency-dependent time lags of 1H~0707--495 could be explained using a dual lamp-post model. They fixed the lower source height at $2~r_{\rm g}$ and found the upper coronal height varying at $\sim 3$--$20~r_{\rm g}$. Our finding also supports that the 1H~0707--495 corona is dynamic and extended as in IRAS~13224--3809, but with smaller physical distance due to their mass difference. 

Previous studies based on the lag-frequency spectra suggested a compact corona for MCG--6-30-15 that is located at $\sim 3~r_{\rm g}$ \citep{Emmanoulopoulos2014, Epitropakis2016}. The extended corona model has never been adopted to formally fit their lag-frequency spectra before. However, our results conclude that the corona in MCG--6-30-15 should also be varied since we observe the variable Granger-lags as in IRAS~13224--3809 and 1H~0707--495. Interestingly, the averaged lags of MCG--6-30-15 are mostly within $\sim 500$--1500~s, similar to IRAS~13224--3809. The MCG--6-30-15 mass is also $\sim 2 \times 10^{6} M_{\odot}$ \citep{Ponti2012} which is comparable to the IRAS~13224--3809 mass. This further suggests that its size and geometry may be analogous to that of IRAS~13224--3809.

In the case of IZw1, the Fe-K reverberation lags were probed using the lag-energy spectra, but no Fe-K reverberation feature was detected \citep{Kara2016}. Nevertheless, \cite{Wilkins2021} applied a different method to the standard Fourier technique and detected the Fe-K lag amplitude of $\sim 746 \pm{157}$~s due to the X-ray flares occurring near the black hole. Here, the evidence of the X-ray reverberation lags in IZw1 is also revealed but in the Fe-L band, with the lags varying mostly within $\lesssim 1000$~s. The IZw1 mass is approximately $2.5 \times 10^7 M_{\odot}$ \citep{Gonzalez2012}, so its corona is more compact than the IRAS~13224--3809 corona. The amplitudes as well as the trends of the lags of Mrk~704 and Mrk~1040 are comparable to what seen in IRAS~13224--3809. Keeping in mind that the mass of Mrk~704 and Mrk~1040 is $1.3 \times 10^{8} M_{\odot}$ \citep{Ponti2012} and $4.0 \times 10^{7} M_{\odot}$ \citep{Gonzalez2012}, respectively, we can infer their relatively more compact corona compared to IRAS~13224--3809.

It is important to note that the GC method is a purely hypothetical test that is physical-model independent. The GC test by itself does not provide an explanation of true causality, so it does not provide information of the origin of these lags. Here, we perform theoretical simulations to show that these lags can possibly be produced by reverberation. The results reveal that the lags are variable in a complex way. The main parameter driving the variable lags might be intrinsic to the corona itself. The size and geometry of the corona may vary progressively even within one individual observation. All of these results support a dynamic corona that, sometimes, can cluster in more than one zone, resulting in the separate competing lags. Using more than two light curves would help reveal more additional hidden factors influencing the target light curve.

The GC technique is a statistical test so justifying the errors of the lags is not straightforward. In fact, the AGN light curves points can be correlated to past values, so the time-domain statistics themselves can be correlated. Therefore, one cannot easily distinguish if it is the light curves themselves being variable or it is the variations of the intrinsic lags that mainly produce the variation of the Granger lags (i.e. the lags from reverberation are indistinguishable from one that happens by chance). It could be useful in the future to take into account the GC test results in conjunction with some of the statistical requirements in order to develop a robust physical model behind the test. Here, the meaning and significance of the lags rely on a given interpretation specific to the reverberation framework. Therefore, the observed variable-lags obtained by the GC test should be thought in the way that they all have a possibility to be produced by reverberation. The results from this technique still needed to be considered together with prior knowledge and known information from Fourier-based time-lags to draw a complete conclusion. Also, the GC technique provides a single number of the lag, hence the information about the structure of the cross-correlation between two light curves is limited. Improving the GC technique itself might help evade this issue but this is beyond the scope of this work. Last but not least, we note that the specific implementation details of the GC test as well as the sliding window algorithm can be further advanced by, e.g., applying it to directly fit the data or using a variable window-size specifically for any targeted segment of the light curve. New and improved statistical analysis may provide us a better way to robustly detect the significance of the lag.

\section*{Acknowledgements}

This research has received funding support from Suranaree University of Technology (SUT) and the NSRF via the Program Management Unit for Human Resources \& Institutional Development, Research and Innovation (PMU-B) (grant number B13F660067). PC thanks for the financial support from SUT (grant number 179349). We thank the anonymous referee for the comments which improved the paper.

\section*{Data availability}
The AGN data can be accessed from {\it XMM-Newton} Observatory (\url{http://nxsa.esac.esa.int}). The modules for the Granger causality test is adopted from {\tt statsmodels.tsa.stattools} available in \url{https://www.statsmodels.org}. The derived data and models underlying this article will be shared on reasonable request to the corresponding author.

\bibliographystyle{mnras}
\bibliography{ecm}


\appendix
\section{Effects of binning}
\label{sec:a1}

\begin{figure*}
\centerline{
\includegraphics*[width=0.5\textwidth]{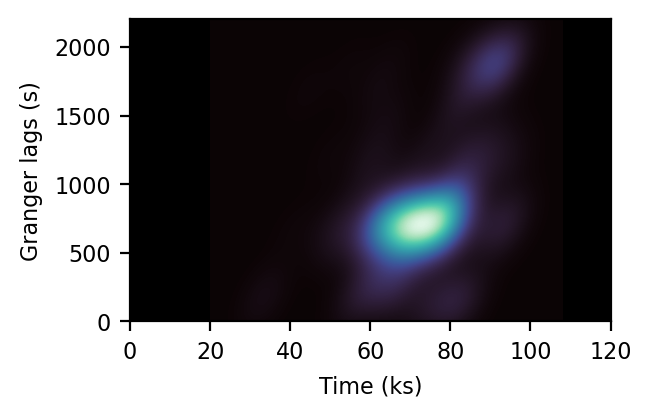}
\put(-195,140){\textcolor{white}{$\Delta t_{i}$ = 50~s, $\Delta t_{f}$ = 250~s, $\delta t_{\rm step}$ = 10~s}}
\includegraphics*[width=0.5\textwidth]{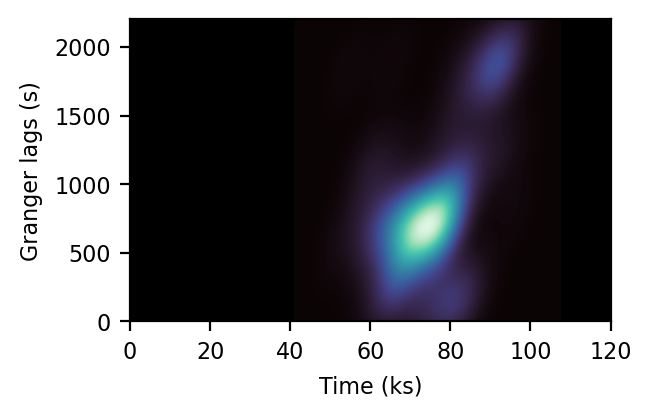}
\put(-195,140){\textcolor{white}{$\Delta t_{i}$ = 50~s, $\Delta t_{f}$ = 250~s, $\delta t_{\rm step}$ = 20~s}}
\vspace{-0.15cm}
}
\centerline{
\includegraphics[width=0.5\textwidth]{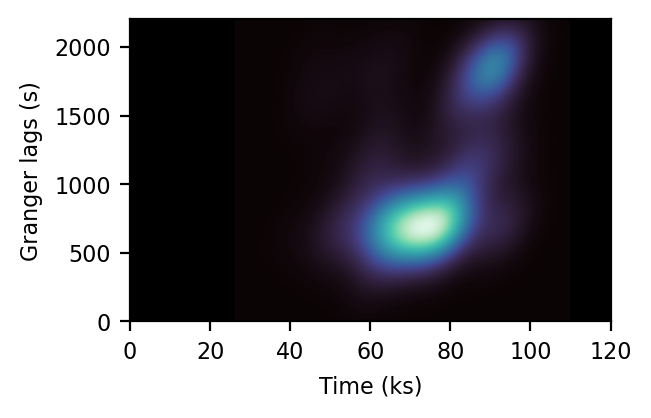}
\put(-195,140){\textcolor{white}{$\Delta t_{i}$ = 100~s, $\Delta t_{f}$ = 300~s, $\delta t_{\rm step}$ = 10~s}}
\includegraphics[width=0.5\textwidth]{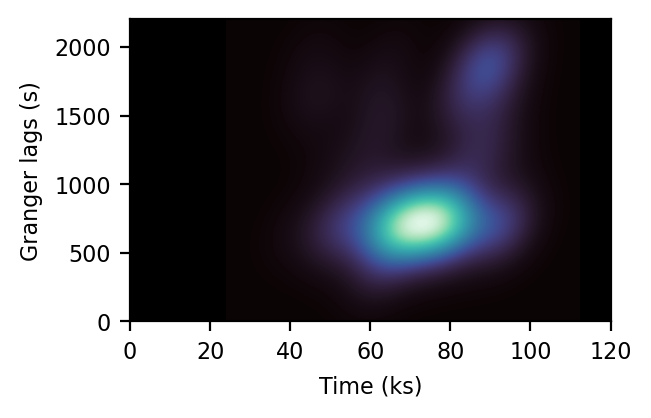}
\put(-195,140){\textcolor{white}{$\Delta t_{i}$ = 100~s, $\Delta t_{f}$ = 300~s, $\delta t_{\rm step}$ = 20~s}}
\vspace{-0.2cm}
}
\caption{Density maps showing the distribution of the Granger-lags detected at a significance level of $p<0.01$ along the light curve of IRAS~13224--3809 rev. no. 2131. Note that the units on the density are arbitrary and higher probability-density values are associated with lighter colors. Top panels: Bin size of the light curve is varied between $\Delta t_{i}= 50$~s and $\Delta t_{f} = 250$~s, with the step size of $\delta t_{\rm step}=10$~s (left panel) and 20~s (right panel). Bottom panels: Bin size is varied between $\Delta t_{i}= 100$~s and $\Delta t_{f} = 300$~s, with the step size of $\delta t_{\rm step}=10$~s (left panel) and 20~s (right panel). Adjusting the bin sizes used in the investigation may affect the Granger-lag profiles, but the trend of how the evolving lags appear as well as the cluster of the lag confidence do not significantly change.}
\label{fig-bin-size}
\end{figure*}

How the bin size affects the distribution of the Granger-lags found along the light curve of IRAS~13224--3809 rev. no. 2131 is shown, as an example, in Fig.~\ref{fig-bin-size}. While there is an evolution of the X-ray reverberation lags within this individual observation, the lags are seen only in the partial segment of the second-half light curve, regardless of the bin and the step size. We see hints of increasing lags within this partial segment. The Granger-lag profiles may change if the bin sizes are changed, however neither the cluster of the lag confidence nor the trend of the evolving lags are considerably altered.

\onecolumn

\end{document}